\documentclass[12pt,preprint]{aastex}
\usepackage{lscape}

\newcommand{\ltsima}{$\; \buildrel < \over \sim \;$}
\newcommand{\simlt}{\lower.5ex\hbox{\ltsima}} 
\newcommand{\gtsima}{$\; \buildrel > \over \sim \;$}
\newcommand{\simgt}{\lower.5ex\hbox{\gtsima}} 

\newcommand{\pedix}[2]{$#1_{\,\mbox{\scriptsize #2}}$}

\newcommand{\feka}{\mbox{Fe I K$\alpha$}}
\newcommand{\fekb}{\mbox{Fe I K$\beta$}}
\newcommand{\feLa}{\mbox{Fe XXVI Ly$\alpha$}}
\newcommand{\nika}{\mbox{Ni K$\alpha$}}
\newcommand{\xmm}{{XMM-\emph{Newton} }}
\newcommand{\bfxmm}{{XMM-{\it \textbf{Newton}} }}
\newcommand{\asca}{{\emph{ASCA} }}
\newcommand{\lum}{erg~s$^{-1}$}
\newcommand{\flux}{{erg~cm$^{-2}$~s$^{-1}$ }}
\newcommand{\nh}{cm$^{-2}$}
\newcommand{\nhsym}{N_{\mbox{\scriptsize H}}}
\renewcommand{\arcsec}{\mbox{$^{\prime\prime}$} }
\newcommand{\kev}{\,\mbox{\scriptsize keV}}

\newcommand{\norm}{photons~keV$^{-1}$~cm$^{-2}$~s$^{-1}$}

\newcommand{\sorg}{AXJ0447-0627}
\newcommand{\sorgs}{AXJ0447-0627 }
\newcommand{\chandra}{{\emph{Chandra} }}

\received{}
\begin{document}

\title{The \bfxmm view of the relativistic spectral features in \sorg}

\author{R. Della Ceca$^1$, L. Ballo$^2$, V. Braito$^1$, and T. Maccacaro$^1$}

\affil{$^1$ INAF - Osservatorio Astronomico di Brera, via Brera 28, 20121 Milan, Italy 
(rdc@brera.mi.astro.it, braito@brera.mi.astro.it, tommaso@brera.mi.astro.it)}
\affil{$^2$ SISSA/ISAS, International School for Advanced Studies, 
            via Beirut 4, 34014 Trieste, Italy (ballo@sissa.it)}

\begin{abstract}

The \xmm observation of the optically Type 1 AGN \sorgs ($z=0.214$)
unambiguously reveals a complex, bright and prominent set of lines  in the 
$4-8\,\:$keV rest frame energy range. 
Although, from a phenomenological point of view, the observed properties can be 
described by a simple power law model plus 5 narrow Gaussian lines 
(at rest frame energies of $\sim 4.49$, $\sim 5.55$, $\sim 6.39$, $\sim 7.02$ 
and $\sim 7.85\:\,$keV),  
we find that a model comprising a power law ($\Gamma \sim 2.2$),  a
reflected relativistic continuum,  a narrow {\feka} line from neutral material
as well as a broad Fe  K$\alpha$ 
relativistic line from a ionized accretion disk  
represents a good physical description of the data.   
The ``double horned'' profile  of the relativistic line implies  an inclination
of the accretion disk of $\sim 45^{\circ}$, and an origin in a narrow 
region of the disk, from \pedix{R}{in}$\sim 19\:\,GM/c^{2}$ to 
\pedix{R}{out}$\sim 30\:\,GM/c^{2}$. The  narrow {\feka} line from neutral material 
is probably produced far from the central black hole, most likely in the 
putative molecular torus. Although some of these
properties have been already found in other 
Type 1 AGN and discussed in the literature, at odd with the objects
reported so far we measure high  equivalent widths (EWs) of the observed lines: 
$\sim 1.4\,\:$keV for the  ``double horned'' relativistic line and $\sim
0.4\,\:$keV  for the narrow  line.

\end{abstract}


\section{Introduction}

Recent \xmm  and \chandra observations of a number of Active Galactic Nuclei have 
discovered  complex spectral features (broad and/or narrow) red-ward
of the well known  fluorescent Fe emission line(s) at $6.4 - 7\:\,$keV
(\citealt{turner02}; \citealt{guainazzi03}; \citealt{yaqoob03}; \citealt{turner3516}; 
\citealt{porquet04}; \citealt{bianchi04};
\citealt{mckernan04}; \citealt{gallo04}).
These features are probably  produced  in (or very close to) the supposed
accretion disk and, as such,  their study  provides primary information 
about the dynamics and  physical processes which are taking  place in the 
innermost part of the AGN (see \citealt{fabian00} and \citealt{reynolds03} 
for a review). 
Among the models invoked to explain these features there are:
{\it i)} localized hot spot on the accretion disk surface due to 
illumination by local flares (e.g. \citealt{dovciak04});
{\it ii)} reprocessed emission from narrow annuli on the surface of the 
accretion disk \citep{gallo04};
{\it iii)} inflow or outflow of material (e.g. the ejected blob model 
proposed by \citealt{turner766} to explain the properties of the narrow lines 
observed in MKN~766); 
and 
{\it iv)} destruction (spallation) by  energetic protons on the accretion disk surface 
of Fe into lower Z elements (mainly Cr and Mn), implying an enhancement 
of the line emission expected from elements of lower abundance \citep{skibo97}.
  
In this paper we use \xmm data to discuss the observed line properties of \sorg, a
broad line AGN at $z=0.214$, discovered during the optical identification
process of the X-ray sources of the \asca ($2-10\:\,$keV) Hard Serendipitous
Survey (\citealt{cagnoni98}; \citealt{rdc99}). 
We assume $H_0=70$~km~s$^{-1}$~Mpc$^{-1}$, 
$\Omega_{\lambda} = 0.7$ and $\Omega_{M} = 0.3$. 

\section{Observations and data reduction}

\subsection{X-ray observations}

\sorgs was detected with a $S/N=5.40$ in the \asca field pointed at NGC~1667
(\asca Sequence~ID$=71032000$) at the nominal \asca position of
$\mbox{RA}$ = 04:47:48.6,  $\mbox{Dec}$ = $-$06:27:50.8 (J2000.0). The net \asca
GIS2 counts from the source are $54 \pm  10$ ($2-10\:\,$keV), corresponding to
a $f_{2-10\:\,\kev} \sim 3 \times 10^{-13}\:\,$\flux (assuming a 
power-law photon
index $\Gamma=1.7$).
The source attracted our attention because of its position in the hardness
ratio diagram (cf. \citealt{rdc99}), indicative of a
hard,  presumably obscured, X-ray source ($HR1=-0.40\pm 0.15$; $HR2=0.32\pm
0.17$).

We thus observed \sorgs with \xmm on Sept~8,~2002 for a total of about
$29\,$ks. The three EPIC cameras (MOS1, MOS2 and pn) 
were operating in full frame mode and with the thin
filter applied. The \xmm data have been cleaned and processed using the 
Science Analysis Software (SAS version~5.4), and analyzed using standard
software packages (FTOOLS version~4.2, XSPEC version~11.2). 
Event files have been thus filtered for high-background time intervals and only
events corresponding to patterns $0 - 12$  (MOS 1\&2) and $0 - 4$ (pn) have been
used (see \citealt{xmmhb}); the net exposure times at
the source position after data cleaning are $\sim 21.4\:\,$ks (MOS1, MOS2)
and $\sim 17.5\:\,$ks (pn).

The \xmm MOS1, MOS2 and pn images in the $0.5-10\:\,$keV energy range reveal a
high signal-to-noise ratio ($S/N \sim 30$ and $\sim 50$ in the MOS and the pn,
respectively) point-like source within the \asca 90\% positional 
error circle ($\sim 2^{\prime}$
radius) of \sorg. This is the
only detected and visible X-ray source in the \asca error circle; the X-ray
position derived using the \xmm data is $\mbox{RA}$ = 04:47:48.62, $\mbox{Dec}$ =
$-$06:28:12 ($\sim 21\arcsec$ away from the nominal \asca position).

Source counts were extracted from a circular region of radius $22.5\arcsec$ for
the MOS and $17.5\arcsec$ for the pn (this smaller radius for the pn is due to
the proximity of a CCDs gap). Background counts were extracted from a nearby
source-free circular region of $\sim 42\arcsec - 50\arcsec$ radius. The net
count rates ($0.5 - 10\:\,$keV energy range) are $0.051\pm
0.002\:\,$counts~s$^{-1}$, $0.054\pm 0.002\:\,$counts~s$^{-1}$ and $0.171\pm
0.004\:\,$counts~s$^{-1}$ for MOS1, MOS2 and pn, respectively; 
the source counts
represents
about  85\% of the total counts in the source extraction region.  No
statistically significant source variability has been detected during the \xmm
observation.  To improve statistics, the MOS1 and MOS2 data have been combined
together, and the MOS and pn spectra have been fitted simultaneously, keeping
the relative normalization free. Source counts were binned so as to have at 
least $20$ counts in each energy bin.  We have also generated our own spectral
response matrices at the source position using the SAS tasks \emph{arfgen} and
\emph{rmfgen}. All the models discussed here have been filtered through the
Galactic absorption column density along the line of sight (\pedix{N}{H,
Gal}$=5.6\times10^{20}\:\,$\nh; \citealt{nh}).
Unless  otherwise stated, fit parameters are quoted in the
rest-frame of \sorgs ($z=0.214$, see below), 
while the figures and the EWs are in the observer frame.

\subsection{Source identification and optical spectroscopy}

A  bright  ($\mbox{RA}$ = 04:47:48.5, $\mbox{Dec}$ = $-$06:28:13; APM red
magnitude = 17.7)  optical source lies about $2\arcsec$ from  the X-ray position
derived using the \xmm data. This object was observed spectroscopically at the
TNG  on  October 5th,~2002. The optical spectrum (not reported here) covers the
wavelength range $\sim 3500-8000\:\,$\AA ~(dispersion of $2.8\:\,$\AA/pixel)
and  clearly shows broad (FWHM $> 6000\:\,$km~s$^{-1}$) MgII, H$\beta$ and
H$\alpha$ lines  as well as narrow (FWHM $< 1000\:\,$km~s$^{-1}$)
[OIII]4959,5007~\AA $\ $ lines.  The optical line properties
and position allow us to classify \sorgs as a  classical broad line AGN  at
$z=0.214 \pm 0.001$. \sorgs also belongs to the \asca Medium Sensitivity Survey,
 so an
independent confirmation of the redshift and of the optical spectral
classification comes from the work presented in \citet{idOpt}.  

\section{X-ray spectral analysis}

A single absorbed power law (PL) model is not a good description of the overall
($0.5 - 10\:\,$keV) spectrum of \sorgs ($\Gamma=2.18 \pm 0.05$; $\nhsym$
consistent with zero; $\chi^2/\mbox{d.o.f}=280.3/250$), since a very large 
discrepancy is present above $4\,\:$keV. 

In Fig.~\ref{fig:ratiopl} we show the ratio between the best fit PL model
(obtained considering only the line-free region from $\sim 0.8$ to $\sim
3\:\,$keV) and the data in the  $2-10\,\:$keV energy  range.  The residuals
show the presence of a  possible  ``line-like'' feature at  $E \sim
3.5\,\:$keV (observer frame), and of a complex structure (several different
lines?) in the  energy range between $\sim 4.5$ and $\sim 7\,\:$keV (observer
frame).  Splitting the total observation into two intervals of similar
exposure times, we do not find convincing evidence of temporal variability of
such structures.

We have been unable to reproduce the complex lines  structure with  a pure
reflected continuum, either normal (PEXRAV model in XSPEC; see \citealt{pex}) 
or relativistic (REFSCH model; see \citealt{pex}; \citealt{disk}).  A  pure
reflected continuum with associated emission lines from Ca, Cr, Fe and Ni (see
e.g. the modeling of the Seyfert~2 galaxy NGC~6552 by \citealt{reynolds94}) is
also unable to reproduce the observed structure since the relative abundances
of the  above elements are different from the expected ones. We also note that
a pure reflected continuum, more typical of optical Type~2 AGNs, is at
odd with the optical spectral classification of \sorg.

To investigate the presence of line-like features and to locate their  energy
centroid we slid a narrow ($\sigma=0.1\,\:$keV) Gaussian template  across the
data between  $4$ and $8\,\:$keV (rest frame), looking for an improvement  to
the fit with respect to the simple PL model.  The results are shown in the
inset of  Fig.~\ref{fig:ratiopl}, where we report the change in fit  statistics
($\Delta \chi^2$) as a function of the centroid  energy position of the narrow
Gaussian line.     This analysis points out the presence of a number of
possible narrow lines, with rest frame 
energy centroids at about $4.5$, $5.6$, $6.4$, $7.0$
and $7.9\,\:$keV (in the last case, the line centroid is not well
constrained).    Note that in the resulting fit statistics a $\Delta \chi^2$
greater than $10.30$ ($5.25$) represents a feature that is significant at more 
than 99\% (90\%) confidence level, so the lines at energies $\sim 4.5$ keV and at 
$\sim 7.9$ keV
are of lower statistical significance with respect to those at 
$\sim 5.6\,\:$keV, $\sim 6.4\,\:$keV and $\sim 7.0\,\:$keV.  Following these indications, we
tried to reproduce the observed spectrum with a power law and five narrow
Gaussian lines.  The best fit spectral parameters are reported in
Table~\ref{tab:gauss} while the ratio between the data and this possible  best 
fit model is shown in  Fig.~\ref{fig:ratiocfr} (panel~a).  

The  line at $6.39^{+0.07}_{-0.06}\,\:$keV  (EW$\,\:\sim  700\,\:$eV) 
is positionally consistent with the {\feka} emission line, while 
the line at $7.02^{+0.29}_{-0.12}\,\:$keV (EW$\,\:\sim  600\,\:$eV)
is positionally consistent
both with the {\fekb} emission line (rest frame energy 
$E=7.058\,\:$keV)
or with the {\feLa} emission line (rest frame energy $E=6.96\,\:$keV).
However the association with the  {\fekb}  line is 
unlikely since the measured flux from this line 
should be at a fixed ratio ($\sim 0.11$) 
with the  {\feka} emission flux, clearly in disagreement 
with the measured EWs.
The association with {\feLa} seems to be  more plausible given that this line 
could be very prominent (and sometimes with an EW comparable with the 
narrow {\feka} line) in Type 1 AGN (see e.g. the case of the Seyfert 1 galaxy 
NGC 7314 discussed in \citealt{yaqoob03}).
The line at 
$7.85^{+0.73}_{-0.76}\,\:$keV is positionally consistent with the \nika, while
for the remaining two lines ($E=4.49^{+0.13}_{-0.17}\:\,$keV and
$5.55^{+0.06}_{-0.05}\:\,$keV) there are no clear associations
with well known and expected elements. The strongest expected lines in the
spallation model are the Cr~K$\alpha$ at 5.4 keV and the Mn~K$\alpha$ at 5.9
keV \citep{skibo97}. Both these lines are ruled out by the mismatch with the
measured energy lines centroid. So, unless an energy shift occurs (but we do
not observe any energy shift for the {\feka} line), the spallation model is an
unlikely explanation of the \xmm data. 
We have also evaluated the upper limits for Fe~XXV~(f) at $E \sim 6.64\:\,$keV
and for Fe~XXV~(r) at $E \sim 6.70\:\,$keV since the strenght of these lines, 
when combined with the strenght of other ionized 
Fe lines, can constrain emission 
models (cfr. \citealt{yaqoob04} and reference therein). 
These two lines are not required by the current data set and the 
90\% upper limit on their EW is $\sim 400\:\,$eV.


The complex structure detected in the spectrum of \sorgs could suggest
a profile of a Fe line produced by an accretion disk. 
We explored this interesting possibility using 
the DISKLINE model \citep{disk}, which assumes 
a non-rotating Schwarzschild black hole\footnote{We 
have also tried the LAOR model \citep{laor}, 
in which the black hole is maximally 
rotating, but because of the limited source
count  statistics we  could not discriminate between the
DISKLINE and LAOR model. 
Since the black hole spin parameter is clearly an over-parameterization 
of the present data set we report here only the results obtained applying the DISKLINE
model. We note however that similar results have been obtained using the 
LAOR model.}.
The relativistic effects have been introduced also in the
description of the reflected continuum, replacing the simple power law model 
with the XSPEC model REFSCH\footnote{
This model is the sum of an e-folded power law primary spectrum 
plus its reflected  component from a ionized relativistic accretion disk.}.

We started the analysis using a model composed of the REFSCH  model 
plus a {\feka} disk line fixing its energy position 
to $6.4\:\,$keV rest-frame; 
since in \sorgs the observed 
lines seems to have a 
significantly larger 
EW than usual we have also added the corresponding 
{\fekb} disk line emission fixing the {\fekb}/{\feka}
ratio to that expected form the theory ($\sim 0.11$).
The best fit spectral
parameters are reported in Table~\ref{tab:relat} while the ratio between the data
and the best fit model is shown in  Fig.~\ref{fig:ratiocfr} (panel~b).
Although the overall fit is statistically acceptable ($\chi_{\nu}^2 = 1.00$), 
the ratio in Fig.~\ref{fig:ratiocfr} (panel~b) shows a 
``line-like'' residual at an 
observed  energy of 
$\sim 6\,\:$keV  ($\sim 7.3\,\:$keV rest frame) that we were unable to reduce. 
We tried to
consider also disk lines associated to
{\nika} at $\sim 7.5\,\:$keV or to {\feLa} at $\sim 6.96\,\:$keV
emission, and/or
allowed the abundance of {\fekb}, {\nika} and {\feLa}
to be a free parameter of the fit, but
we were  unable to take into account such ``line-like'' structure. 

Thus we tried the same  model first used by \citet{weaver97} 
to describe the spectral  properties of  MCG-5-23-16:  a narrow
{\feka} component (\pedix{E}{line} fixed at  $6.4\,\:$keV)
plus  a  broad relativistic line  component (DISKLINE model) along with the
underlying continuum (REFSCH model). The best fit spectral parameters   are
reported in Table~\ref{tab:relat}, the ratio between the data  and the best fit
model is shown in  Fig.~\ref{fig:ratiocfr} (panel~c), while the folded model 
and the model itself are reported in  Fig.~\ref{fig:foldedrel}. The observed
line position ($E=6.61\pm 0.11\:\,$keV) of the broad relativistic line  is  
inconsistent with the {\feka}
from neutral material but  strongly suggests that it is due to  
Fe K$\alpha$ 
emission from  ionized He-like material. Overall this modeling provide a good
description  of the broad band spectral properties of \sorg. 
We have also tried to add a relativistic emission line from 
{\feLa} to the best 
fit model reported in Table 2; such line can be accommodated 
within the present data set 
(with an EW of $\sim 15\:\,$eV) but it is not statistically required. 
The observed flux
and the intrinsic luminosity in the $0.5-10\,\;$keV energy range are $(6.5\pm
0.4) \times10^{-13}\,\:$\flux and  $(8.9\pm 0.5) \times10^{43}\,\:$\lum, the
errors reflecting the uncertainties  on the best fit model. 
We note that the $2-10\,\:$keV flux measured with  \xmm ($\sim 3.6 \times
10^{-13}\,\:$erg~cm$^{-2}$~s$^{-1}$) is in very good agreement with the $2-10\,\:$keV flux
measured with  \emph{ASCA}.

\section{Discussion and Conclusions}

Using \xmm data we have revealed that the optically Type 1 AGN \sorgs at
$z=0.214$  is characterized  by a complex of bright and prominent set of lines
in the  $4.5 - 8.0\:\,$keV energy range (rest frame).

We have shown that these lines can be reasonably well  reproduced by a physical 
model comprising a power law, a reflected relativistic continuum, a narrow
{\feka} line from neutral material and a  Fe He-like 
K$\alpha$  relativistic line  from a
ionized accretion disk.   Although not well constrained, the  best fit ionization
parameter ($\xi$ in  Table~\ref{tab:relat}) is consistent with the  Fe
ionization state, as deduced from the best fit line position (cf.
\citealt{matt93}). 
A similar modeling of the Fe line properties (a narrow plus a broad  relativistic
component) has been found to describe, for example,  the spectral properties of
the Seyfert~1.9 MCG-5-23-16 \citep{weaver97}, of the  radio-quiet quasar
MKN~205  \citep{reeves01} and of the Seyfert~1 NGC~3516 \citep{turner02};  the
presence of the Fe relativistic line  from a high ionized accretion
disk has been unambiguously reported  in the case of MKN~205. As already discussed by
the above authors, the most likely origin of the  narrow $6.4\:\,$keV  component is from
neutral matter distant from the black hole, e.g.  the putative molecular torus.

The resulting best fit of the relativistic ``double horned''  profile implies
an inclination of the accretion disk of $\sim 45^{\circ}$, as well as that the
observed lines should be produced in a narrow region of the disk from 
\pedix{R}{in}$\sim 19\:\,GM/c^{2}$ to \pedix{R}{out}$\sim 30\:\,GM/c^{2}$.  As in the
case of e.g. ESO~198-G24 \citep{guainazzi03} and  NAB~0205+024
\citep{gallo04},  a few alternative possibilities can be conjectured to explain
why the inner radius is larger than the last stable orbit.   The first
possibility is that the disk is highly ionized in the inner part, so most of 
the Fe is completely stripped off and the production of the Fe lines is
suppressed. Second, the accretion disk in \sorgs could be  truncated at 
$10 - 15\ GM/c^{2}$ (see \citealt{muller04}). 
Finally the relativistic line can be
produced by a localized hot spot on the accretion disk surface 
(e.g. \citealt{dovciak04}). 

All the properties discussed above have been also observed in other AGNs  and
seem to be in agreement with the expectation from accretion disk theory.
However  there is an observed property that is very unusual in \sorgs  and
makes this object unique: the very  large  EW observed, which
is at least a factor 5  greater than that usually measured in other Type 1 AGNs 
(see e.g. \citealt{yaqoob04} and reference therein) or
expected from an accretion disk around a Schwarzschild black hole 
(\citealt{matt92}; \citealt{matt93}).
According to the modelling reported  in \cite{martocchia96} and 
\cite{miniutti04}, a high EW could be explained if the primary 
X-ray source (illuminating both the observer and the accretion disk)
is located very close to a central and maximally rotating Kerr black hole. 
However such a combination should also imply a very high value of the 
reflection parameter $R$ and a reflection dominated source, probably 
in disagreement  with the best fit found here.

A way to solve part of these problems is to assume that the emission features appear 
much stronger than normal because the continuum is strongly absorbed.
We have tested this possibility 
 by adding a partial covering absorption model in front to the 
underlying continuum. Unless the primary AGN emission
is heavly absorbed ($\nhsym \sim 10^{25}\:\,$cm$^{-2}$) and therefore with signatures
falling outside the \xmm bandpass (but this is clearly at odd 
with the optical spectral classification of \sorg), the best fit absorbing 
$\nhsym$ and covering fraction ($\sim 9\times 10^{21}\:\,$cm$^{-2}$ and $\sim 0.2$, 
respectively) imply that absorption effects can not take into account 
the strong emission features observed.

This said, and with the caveat that the model proposed here could be not
fully  appropriate 
(e.g. we have already pointed out that a phenomenological description of the 
data can be also obtained by a simple  power law model 
plus 5 narrow Gaussian lines at rest frame energies of 
4.49$^{+0.13}_{-0.17}\:\,$keV, 
5.55$^{+0.06}_{-0.05}\:\,$keV,
6.39$^{+0.07}_{-0.06}\:\,$keV,
7.02$^{+0.29}_{-0.12}\:\,$keV and
7.85$^{+0.73}_{-0.76}\:\,$keV), 
we would like to note that the emission lines  in \sorgs
have a total EW of $\sim 2\,\:$keV.  This is an observational
result and, as such, is model independent. These lines deserve further
attention and a deeper investigation since any model proposed to describe the
X-ray spectral  properties of \sorgs should be able to explain their large 
EW.

Finally we would like to note that \sorgs was selected as a target  for an \xmm
observation  because of its observed \asca hardness ratios, indicative of a
hard, presumably  obscured, X-ray source (cf. \citealt{rdc99}). The \xmm
observations reported here have  shown that its hard X-ray colors are due
to the strong line complex in the observed $4.5-8.0\,\:$keV energy range rather
than to absorption effects.  There have been many claims in recent years 
about a substantial fraction ($\sim 10$\%) of X-ray absorbed optically classified
Type~1 AGN (with strong implications for AGN unification models and
synthesis  of the cosmic X-ray background) based mainly on poor quality X-ray
data (e.g. hardness ratios; see \citealt{willott04}). The result presented here indicates that a number
of these sources, {\it thought} to be X-ray absorbed Type~1 AGNs on the basis
of their hardness ratios, could instead be X-ray un-absorbed AGNs with
substantial and complex X-ray line emission (see also 
\citealt{macca04}).

\acknowledgements

Based on observations obtained with XMM-\emph{Newton} (an ESA science mission
with instruments and contributions directly funded by ESA Member States and the
USA, NASA) and with the Telescopio Nazionale Galileo (operated in the island of
La Palma by the Centro Galileo Galilei of the INAF in the Spanish Observatorio
del Roque de los Muchacos of the Instituto de Astrofisica de Canarias). 
We would like to thank the anonymous referee for her/his useful comments that 
have contributed improving this paper. 
We
thank M. Cappi, M.  Dadina, A. Caccianiga, L. Maraschi, G. Matt and P. Severgnini 
for useful discussions. 
This work has received partial financial support from the Italian
Space Agency (ASI grant I/R/062/02) and from the MIUR
(Cofin-03-02-23).

\clearpage

 \begin{table*}
  \begin{center}
  \caption{Results of the spectral analysis ($0.5 - 10.0\,\:$keV in the observed
  frame) - Power law plus 5~narrow Gaussian lines.}\label{tab:gauss}
  \vspace{1cm}
   \begin{tabular}{ccccccc}
    \hline
    \hline
    \multicolumn{2}{c}{Power Law} & & \multicolumn{3}{c}{Lines} & \\
    \cline{1-2} \cline{4-6}
     $\Gamma$ & Norm        & & \pedix{E}{r.f.} & Norm      & EW & $\chi^2/$d.o.f.\\
     & & & (keV) & & (eV) \\
     (1) & (2) & & (3) & (4) & (5) & (6) \\
    \hline
    \multicolumn{7}{c}{}\\
     2.24$^{+0.06}_{-0.05}$ & 2.38$^{+0.13}_{-0.06}$ & &
                           4.49$^{+0.13}_{-0.17}$ & 0.92$^{+0.71}_{u}$   &  93$^{+72}_{u}$   & 226.3/240\\
    \multicolumn{3}{c}{} & 5.55$^{+0.06}_{-0.05}$ & 2.68$^{+0.94}_{-1.01}$ & 436$^{+152}_{-164}$ & \\
    \multicolumn{3}{c}{} & 6.39$^{+0.07}_{-0.06}$ & 3.13$^{+0.82}_{-1.26}$ & 700$^{+182}_{-290}$ & \\
    \multicolumn{3}{c}{} & 7.02$^{+0.29}_{-0.12}$ & 2.18$^{+0.91}_{-1.08}$ & 602$^{+251}_{-299}$ & \\
    \multicolumn{3}{c}{} & 7.85$^{+0.73}_{-0.76}$ & 0.74$^{+0.94}_{u}$   & 263$^{+331}_{u}$  &
    \vspace{0.2cm}\\
    \hline
    \multicolumn{7}{c}{}\\
    \multicolumn{7}{l}{\footnotesize  Columns are as follow: }\\
    \multicolumn{7}{l}{\footnotesize  Column 1: power law photon index; Column 2: normalization of the power law in  units of }\\
    \multicolumn{7}{l}{\footnotesize $10^{-4}\,$\norm~@~$1\,$keV; Column 3: rest-frame energy centroid of the narrow}\\
    \multicolumn{7}{l}{\footnotesize  Gaussian line; Column 4: normalization in units of $10^{-6}\,$photons~keV$^{-1}$~cm$^{-2}$~s$^{-1}$ in the line;} \\
    \multicolumn{7}{l}{\footnotesize  Column 5: equivalent width of the line; Column 6:  $\chi^2$ and number of degrees of freedom.}
    \vspace{0.2cm} \\
    \multicolumn{7}{l}{\footnotesize  NOTE: Errors are quoted at the 90\% confidence level for 1 parameter of interest ($\Delta \chi ^2=2.71$); }\\
    \multicolumn{7}{l}{\footnotesize  $u$: unconstrained parameter.}\\
  \end{tabular}
  \end{center}
 \end{table*}

\clearpage

\clearpage

\begin{landscape}
 \begin{table*}
 \scriptsize
  \begin{center}
  \caption{Results of the spectral analysis ($0.5 - 10.0\,\:$keV in the observed frame) - REFSCH model +~Fe K$\alpha$ relativistic line +~Fe (relativistic or Gaussian) line.}\label{tab:relat}
  \vspace{1cm}
   \begin{tabular}{cccccccccccc}
    \hline
    \hline
    \multicolumn{7}{c}{REFSCH$^{(a)}$} & & \multicolumn{3}{c}{DISKLINE} & \\
    \multicolumn{7}{c}{} & & \multicolumn{3}{c}{DISKLINE or GAUSSIAN} & \\
    \cline{1-7} \cline{9-11}    
     $\Gamma$ & $i$ & \pedix{R}{in} & \pedix{R}{out} & $\xi$ & $R$ & Norm & & \pedix{E}{r.f.} & Norm & EW$^{(b)}$ & $\chi^2$\\
     & & & & & & & & (keV) & & (keV) & [d.o.f.]\\
     (1) & (2) & (3) & (4) & (5) & (6) & (7) & & (8) & (9) & (10) & (11) 
    \vspace{0.2cm}\\
    \hline
    \hline
    2.26$^{+0.04}_{-0.07}$ & 25.0$^{+4.4}_{-3.0}{(c)}$ & 16.2$^{u}_{-3.9}{(d)}$ & 25.2$^{+13.5}_{u}{(d)}$ & 1$f$  & $0.44-2.00$ & 1.55$\pm 0.05$         & & 6.4$f$  & 5.3$^{+1.6}_{-1.2}{(c)}$ & 1.3$^{+0.4}_{-0.3}$ & 235.94 \\
                           &                                &                               &                                &         \multicolumn{4}{r}{\it relativistic \fekb:}     & 7.06$f$ & 0.6$^{+0.2}_{-0.1}{(c)}$ & 0.18$^{+0.05}_{-0.04}$ & [246] 
    \vspace{0.2cm}\\
    \hline
    2.21$^{+0.09}_{-0.15}$ & 45.1$^{+6.8}_{-7.4}$~${(c)}$ & 19.5$^{u}_{-6.0}$~${(d)}$ & 29.5$^{+102.1}_{u}$ & 576.4$^{+956.6}_{u}$~${(c)}$ & 1.00$^{+1.00}_{-0.95}$ &  1.53$^{+0.06}_{-0.09}$   & & 6.61$^{+0.11}_{-0.12}$~${(c)}$ & 6.2$^{+2.1}_{-2.5}$ & $1.4^{+0.7}_{-0.4}$ & 223.2 \\
                           &                                 &                             &                      &                                  \multicolumn{4}{r}{\it narrow \feka:}              & 6.4$f$                            & 1.7$^{+1.1}_{-1.0}$ & $0.4^{+0.2}_{-0.2}$ & [243]
    \vspace{0.2cm}\\
    \hline
    \multicolumn{12}{c}{}\\
    \multicolumn{12}{l}{\footnotesize  Columns are as follow: }\\
    \multicolumn{12}{l}{\footnotesize  Column 1: power law photon index; Column 2: inclination angle (degrees); Column 3: inner disk radius in units of $GM/c^{2}$;}\\
    \multicolumn{12}{l}{\footnotesize  Column 4: outer disk radius in units of $GM/c^{2}$; Column 5: disk ionization parameter in units of erg~cm~s$^{-1}$;} \\
    \multicolumn{12}{l}{\footnotesize  Column 6: reflection scaling factor; Column 7: photon flux~@~$1\:$keV of the cutoff broken power law only (no reflection) in the observed } \\
    \multicolumn{12}{l}{\footnotesize  frame in units of $10^{-4}\,$\norm; Column 8: rest-frame energy centroid of the line; Column 9: normalization in units } \\
    \multicolumn{12}{l}{\footnotesize  of $10^{-6}\,$photons~cm$^{-2}$~s$^{-1}$ in the line; Column 10: equivalent width of the line; Column 11: $\chi^2$ and number of degrees of freedom.}
    \vspace{0.2cm}\\
    \multicolumn{12}{l}{\footnotesize NOTE: Errors are quoted at the 90\% confidence level for 1 parameter of interest ($\Delta \chi ^2=2.71$). During the fit,}\\ 
    \multicolumn{12}{l}{\footnotesize the disk parameters of the different components were tied together; $f$: fixed parameter; $u$: unconstrained parameter. }\\
    \multicolumn{12}{l}{\footnotesize$^{(a)}$ The following parameters have been fixed during all the fits: cutoff energy \pedix{E}{c}$=100\,\:$keV; disk temperature $T=3\times10^{4}\,\:$K; } \\
    \multicolumn{12}{l}{\footnotesize power law index for reflection emissivity $\beta=-3$.} \\
    \multicolumn{12}{l}{\footnotesize$^{(b)}$ Equivalent widths are computed with respect to the REFSCH underlying continuum.} \\
    \multicolumn{12}{l}{\footnotesize$^{(c)}$ Errors were calculated fixing at the best fit values \pedix{R}{in} and \pedix{R}{out} (inclination of the accretion disk and relativistic line energy); } \\
    \multicolumn{12}{l}{\footnotesize \pedix{R}{in}, \pedix{R}{out} and $\Gamma$ (relativistic line normalization); $R$ and Gaussian line normalization (disk ionization parameter). } \\
    \multicolumn{12}{l}{\footnotesize$^{(d)}$ Errors have been evaluated performing a fit while stepping the value of \pedix{R}{in} (\pedix{R}{out}) through the range $6.0 - 25.0$ ($20.0 - 200.0$). } \\
   \end{tabular}
  \end{center}
 \end{table*}
\end{landscape}

\clearpage

\centerline{ \bf Figure Captions}

\figcaption[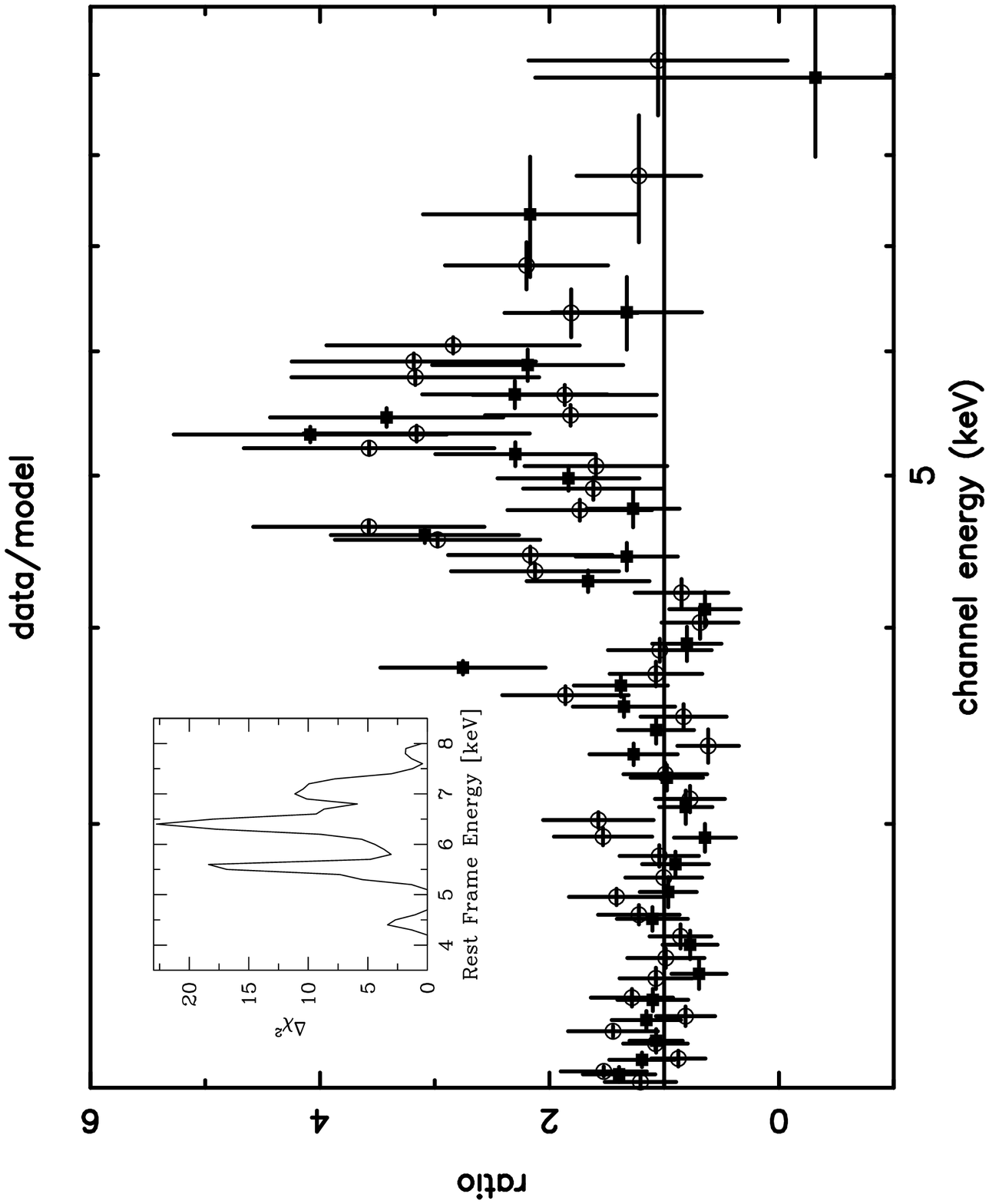]{Ratio between the $2 - 10\:\,$keV MOS (filled squares) and pn (open circles) data
and the best fit power law model (limited to the $\sim 0.8$ to $\sim 3\:\,$keV energy range:
$\Gamma=2.24^{+0.21}_{-0.08}$).  In the inset we report the change
in fit  statistic ($\Delta \chi^2$) as a function of the centroid  energy
position of a narrow Gaussian line model that was stepped across the data; 
the comparison model is the  underlying power law continuum.
\label{fig:ratiopl}}

\figcaption[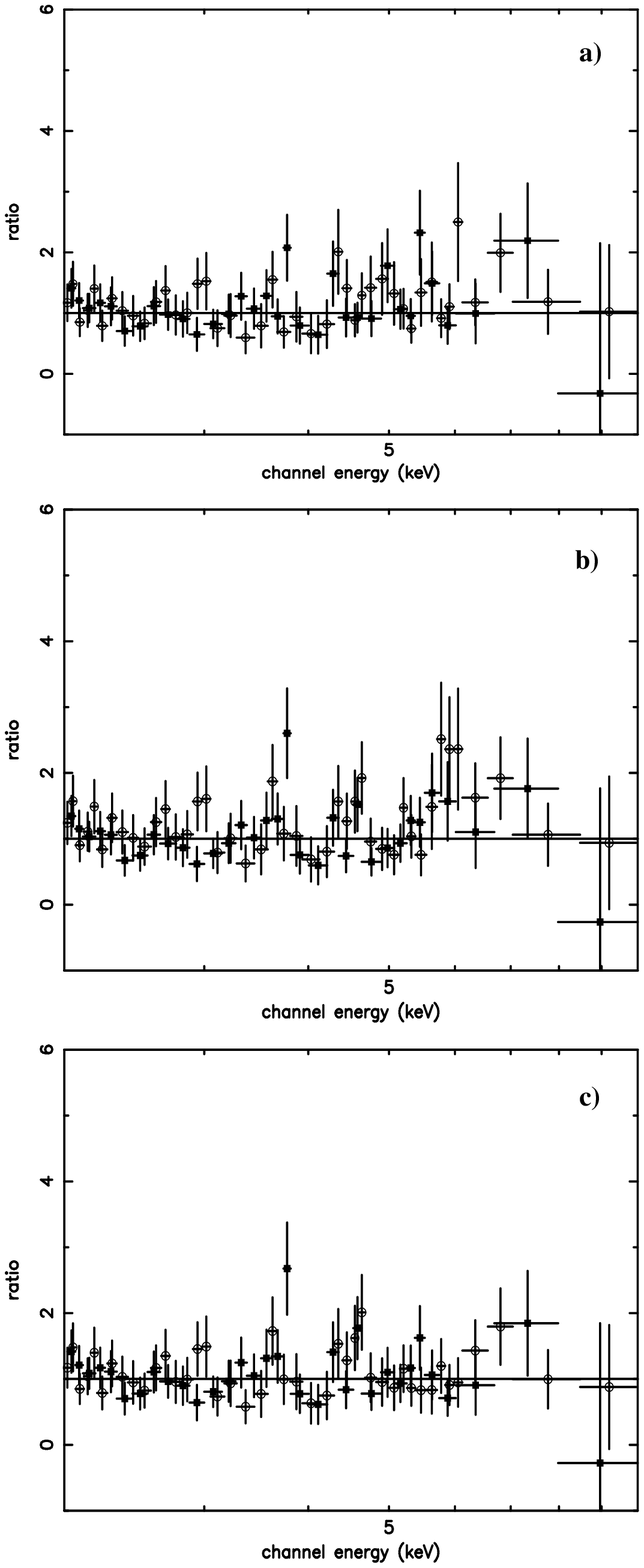]{
{\it Panel~a}: ratio between the $2 - 10\:\,$keV MOS (filled squares) and pn (open circles) 
data and the best fit spectral model composed by a 
power law ($\Gamma=2.24^{+0.06}_{-0.05}$) plus five Gaussian lines as detailed 
in Table~\ref{tab:gauss}.
{\it Panel~b}: ratio between the data and the best fit spectral model composed by 
the REFSCH model  
(an e-folded power law primary spectrum 
plus its reflected  component from a ionized relativistic accretion disk)
plus relativistic {\feka} and {\fekb} lines as detailed in Table~\ref{tab:relat}. 
{\it Panel~c}: ratio between the data and the best fit spectral model composed by
the REFSCH model 
plus a narrow {\feka} line from neutral material and a broad
Fe relativistic line 
from a ionized accretion disk, as 
detailed in Table~\ref{tab:relat}.
\label{fig:ratiocfr}}

\figcaption[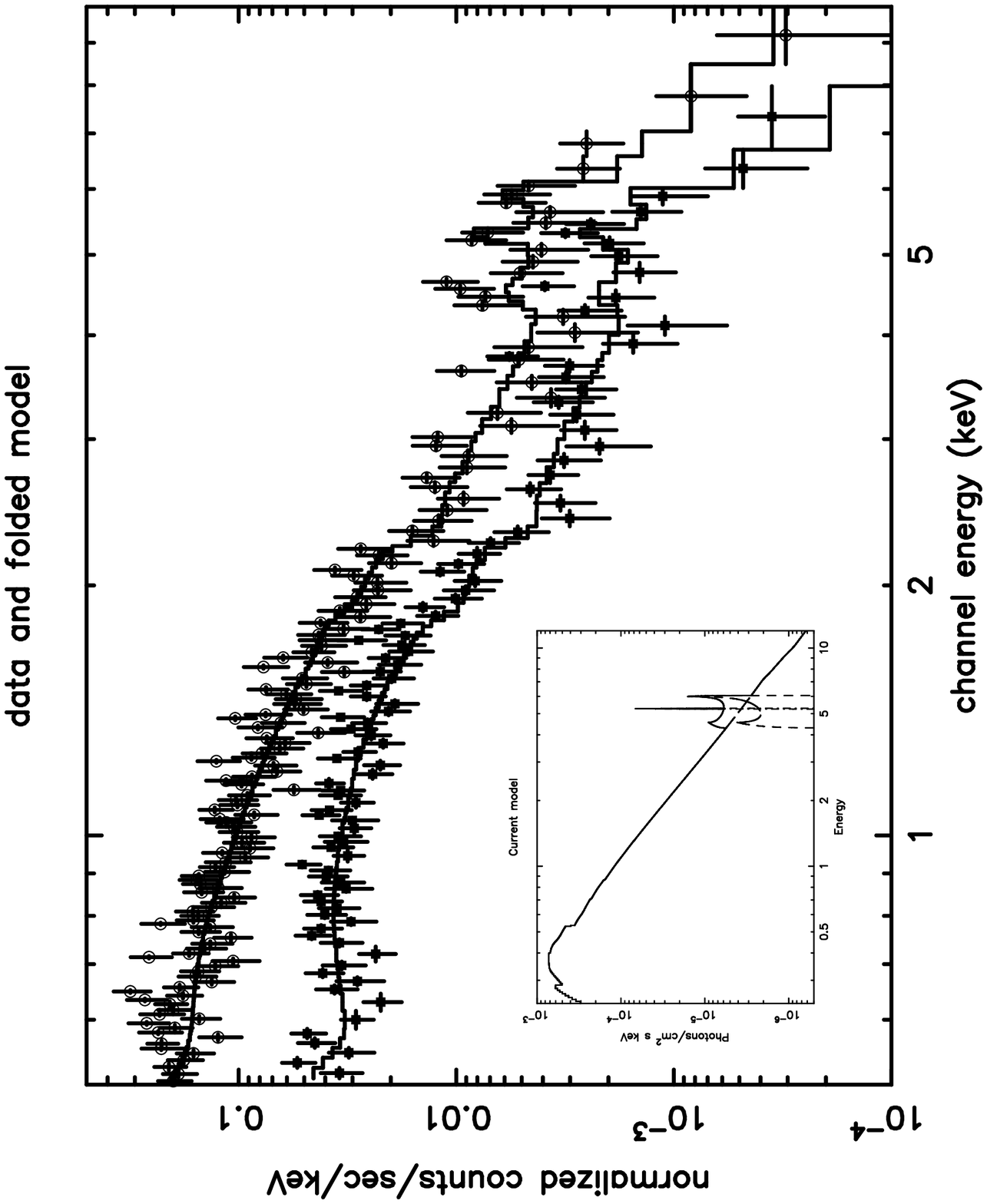]{
MOS (filled squares) and pn (open circles) folded spectra fitted with 
the REFSCH model plus a
narrow {\feka} line from neutral material and a  broad Fe relativistic line 
from a ionized accretion disk.
In the inset we show the best fit model as detailed in 
Table~\ref{tab:relat}.
\label{fig:foldedrel}}

\newpage
\plotone{f1.eps}
\newpage
\plotone{f2.eps}
\newpage
\plotone{f3.eps}

\end{document}